\documentclass[11pt]{revtex4}    
\usepackage{amssymb,epsf}
\usepackage{latexsym}
\begin{document}

\title{ $(n+1)$-Dimensional Lorentzian Wormholes in an Expanding Cosmological Background  }
\author{  E. Ebrahimi
\footnote{email: esmebrahimi@gmail.com} and N. Riazi
\footnote{email: riazi@physics.susc.ac.ir}} \affiliation{Physics
Department and Biruni Observatory, Shiraz University, Shiraz
71454, Iran\\}

\begin{abstract}
We discuss $(n+1)$-dimensional dynamical wormholes in an evolving
cosmological background with a throat expanding with time. These
solutions are examined in the general relativity framework. A
linear relation between diagonal elements of an anisotropic
energy-momentum tensor is used to obtain the solutions. The
energy-momentum tensor elements approach the vacuum case when we
are far from the central object for one class of solutions.
Finally, we discuss the energy-momentum tensor which supports this
geometry, taking into account the energy conditions .
\end{abstract}
\pacs{04.20.Jb, 04.40.Nr, 04.50.+h}
\maketitle
\section{Introduction}
Wormholes are  hypothetical objects which connect two distant
parts of the same  spacetime  or two different spacetimes by a
throat-like object which has the minimum radius of the spacetime.
Although a wormhole solution first entered the physics literature
in 1916 \cite{reff}, the concept was first considered seriously in
1935 by Einstein and Rosen \cite{1} which was later called
Einstein-Rosen bridge but the word wormhole was first time coined
by wheeler \cite{2} in 1957. A more interesting analysis of
wormholes was performed by Morris and Thorne in 1988 and they
presented a new kind of wormhole (traversable wormhole) for the
first time \cite{3}. It was known from before that matter we need
to support such a geometry violates the weak and strong energy
conditions near the throat \cite{4,5}. Morris and Thorne
reconsidered these conditions for a traversable wormhole \cite{3}.
Since the matter that supports this geometry doesn't satisfy the
common energy conditions they called it 'exotic'. An example of
exotic matter is matter with negative energy density \cite{3}.

Another property of these wormholes is the possibility of
transforming them into time machines for backward time traveling
\cite{6,7} and thereby, perhaps for causality violation by closed
timelike curves. Teo \cite{8} found that the null energy condition
($NEC$) is violated by stationary, axially symmetric, traversable
wormholes but there can be classes of geodesics which do not cross
energy condition violating regions. Another case of evolving
wormholes may alter this situation\cite{9,10}. It is known that we
can have violations of weak energy condition ($WEC$) due to some
quantum mechanical effects (such as Casimir effect) \cite{11}. If
we search for this effect in the history of evolving cosmos, we
can find such a situation at the quantum cosmological era when
quantum gravity is dominant. Following the inflation theory by A.
Guth \cite{12} it has been supposed that non-trivial topological
objects such as microscopic wormholes may have been formed during
that era and then enlarged to macroscopic objects with expansion
of the universe \cite{13,14,10,15}. Although most studies have
focused on four dimensions, with the advent of string theory that
demands higher dimensional spacetimes, it is natural to examine
the possibility of wormholes beyond the ordinary four dimensions.
Euclidean wormholes in string theory have recently been studied
\cite{16,17,18,19,20} as solutions of supergravities. With this
motivation, we are going to choose the generalized
Friedmann-Robertson-Walker ($FRW$) metric in $(n+1)$-dimensions
and investigate exact Lorentzian wormhole geometries with
spherically symmetry.

This paper is organized in the following manner: In sec.
\ref{field} we present the ansatz metric and the resulting
solutions in 4-dimensions. In section \ref{ndim}, we extend the
solutions to (n+1)-dimensions. In section \ref{energy}, we
investigate the corresponding energy-momentum tensor and determine
the exoticity parameter. The last section is devoted to
conclusions and closing remarks.

\section{Field Equations\label{field}}

It is very common today that we start studying cosmology with the
so-called 'cosmological principle'. It states that the universe at
large scale is homogeneous and isotropic. With this assumption, we
find out that the metric we need to demonstrate such a spacetime
in 4-dimensions is as follows

\begin{equation}
ds^2= -dt^2+R(t)^2\left[\frac {dr^2}{1
 - kr^2}+ r^2(d\theta ^2+
sin^2\theta \,d\phi ^2)\right], \label{FE1}
\end{equation}
which is known as the Robertson-Walker (RW)  metric \cite{21}. The
coordinate system $(r,\theta,\phi)$ used here is the so-called
'co-moving coordinate system'. $R(t)$ is the scale factor and the
only dynamical parameter to determine. $k=0,\pm1$ correspond to
spatially flat, closed and open spacetimes respectively. This
metric contains a high degree of symmetry which is demonstrated by
its six killing vectors.

For our aim, we need to generalize this metric to
$(n+1)$-dimensions and reduce its symmetry. We break its
homogeneity by replacing $1/(1-kr^2)$ with $1+a(r)$. This metric
is still isotropic about $r=0$ but not necessarily homogeneous. We
therefore write our ansatz metric as
\begin{equation}
ds^2= -  dt^2 + R(t)^2\left[(1 +a(r))\,dr^2 + r ^2 d\Omega_{n-1}
^2\right],  \label{FE2}
\end{equation}
where $a(r)$ is an unknown function. It is clear that the
Robertson-Walker metric is a special case of this metric. With
this ansatz metric, we look at our equations. We start with $n=4$
and then extend the solutions to arbitrary $n$. In the first step
we write our ansatz metric for 5-dimensions (n=4)

\begin{equation}
ds^2= - dt^2+R(t)^2\left[(1 +a(r))\,dr^2 + r
^2(d\theta^2+sin^2\theta \,(d\phi^2+sin^2\phi \, d\psi^2)\right].
\label{FE201}
\end{equation}

The non-vanishing components of the Einstein tensor for our ansatz
metric read

\begin{equation}
 \mathit{G_{t}^{t}} =  - {\displaystyle \frac {6\,
\mathrm{\dot R}^{2}}{\mathrm{R}^{2}}}  - \frac{3}{2}{\displaystyle
\frac { r\,\mathrm{a^{\prime}}+ \mathrm{2a}(1 + \mathrm{a
})}{\mathrm{R}^{2}\,r^{2}\,(1 + \mathrm{a})^{2}}},\label{FE3}
\end{equation}
\begin{equation}
\mathit{G_{r}^{r}}= -3(\frac{R\ddot R+\dot
R^2}{R^2})-3\frac{a}{R^2r^2(1+a)},\label{FE4}
\end{equation}
\begin{equation}
\mathit{G_{\theta}^{\theta}}=-3(\frac{R\ddot R+\dot
R^2}{R^2})-\frac{ra^{\prime}+a(1+a)}{R^2r^2(1+a)^2}, \label{FE5}
\end{equation}
\begin{equation}\label{FE6}
G_{\theta}^{\theta}=G_{\phi}^{\phi}=...,
\end{equation}
in which 'dot' is derivative with respect to $t$, while 'prime' is
derivative with respect to $r$.

Such a geometry is supported by an anisotropic, diagonal
energy-momentum tensor:

\begin{equation}
\rho  = -\frac{1}{8\pi G}T_{t}^{t}= \frac {1}{8\,{\pi
\,G}}\,\left[ {\displaystyle \frac {6\, \mathrm{\dot
R}^{2}}{\mathrm{R}^{2}}}  + \frac{3}{2}{\displaystyle \frac {
r\,\mathrm{a^{\prime}}+ \mathrm{2a}(1 + \mathrm{a
})}{\mathrm{R}^{2}\,r^{2}\,(1 + \mathrm{a})^{2}}}\right],
\label{FE7}
\end{equation}

\begin{equation}
P_{r} =\frac{1}{8\pi G}T_{r}^{r}=\frac{1}{8\pi
G}\left[-3(\frac{R\ddot R+\dot
R^2}{R^2})-3\frac{a}{R^2r^2(1+a)}\right], \label{FE8}
\end{equation}

\begin{equation}
P_{t} =\frac{1}{8\pi G}T_{\theta}^{\theta}=\frac{1}{8\pi
G}\left[-3(\frac{R\ddot R+\dot
R^2}{R^2})-\frac{ra^{\prime}+a(1+a)}{R^2r^2(1+a)^2}\right].\label{FE9}
\end{equation}

In these relations $P_{t}$ and $P_{r}$ are the transverse and
radial pressures, respectively.

Here, we assume the following equation of state

\begin{equation}
\rho  =  - {\displaystyle \frac {\mathit{P_r} + \gamma \,\mathit{
P_t}}{1 + \gamma }}, \label{FE10}
\end{equation}
where $\gamma$ depends on $n$ the dimension of the space. This
equation reduces to the vacuum equation of state $P=-\rho$ when
$P_r=P_t$.

Using equations (\ref{FE7}-\ref{FE9}) and relation (\ref{FE10}) in
5-dimensions (n=4), we obtain:

\begin{eqnarray}
& &\left[ {\displaystyle \frac {6\, \mathrm{\dot
R}^{2}}{\mathrm{R}^{2}}}  + \frac{3}{2}{\displaystyle \frac {
r\,\mathrm{a^{\prime}}+ \mathrm{2a}(1 + \mathrm{a
})}{\mathrm{R}^{2}\,r^{2}\,(1 + \mathrm{a})^{2}}}\right]
+\\\nonumber & & \frac{1}{1+\gamma}\left[ \left[-3(\frac{R\ddot
R+\dot R^2}{R^2})-3\frac{a}{R^2r^2(1+a)}\right] +\gamma
\left[-3(\frac{R\ddot R+\dot
R^2}{R^2})-\frac{ra^{\prime}+a(1+a)}{R^2r^2(1+a)^2}\right]\right]=0.\label{FE11}
\end{eqnarray}

Fortunately, this equation can be separated into radial and
temporal equations.

\begin{equation}
6R\ddot R -6\dot R^2=\frac{(\gamma+3)ra^{\prime}+4\gamma
a(1+a)}{r^2(1+\gamma)(1+a)^2} \label{FE12}
\end{equation}

This equation can be easily solved and has the following
solutions:

\begin{equation}
R(t)=\frac{1}{2}\sqrt{\frac{-1}{C_4}}(e^{(\frac{t}{C_2})}+e^{(-\frac{t}{C_2})}),
 \label{FE13}
\end{equation}

\begin{equation}
1+a(r)= \frac {1}{1+\frac{C_4}{C^2_2}r^{2}+C_1r^{-(\frac
{4\gamma}{3+\gamma})}}. \label{FE14}
\end{equation}

The energy-momentum tensor needed to support this geometry has the
following components:
\begin{equation}
\rho=\frac{3}{4\pi
GC_2^2}-\frac{3r^{-\frac{6(\gamma+1)}{\gamma+3}}C_1(\gamma-3)}{2\pi
GC_4(e^{(\frac{t}{C_2})}+e^{(-\frac{t}{C_2})})^2(\gamma+3)},\label{FE15}
\end{equation}

\begin{equation}
P_r=-\frac{3}{4\pi
GC_2^2}-\frac{3r^{-\frac{6(\gamma+1)}{\gamma+3}}C_1}{2\pi
GC_4(e^{(\frac{t}{C_2})}+e^{(-\frac{t}{C_2})})^2},\label{FE16}
\end{equation}
and
\begin{equation}
P_t=-\frac{3}{4\pi
GC_2^2}+\frac{3r^{-\frac{6(\gamma+1)}{\gamma+3}}C_1(\gamma-1)}{2\pi
GC_4(e^{(\frac{t}{C_2})}+e^{(-\frac{t}{C_2})})^2(\gamma+3)}.\label{FE17}
\end{equation}

\section{$(n+1)$-Dimensional solutions\label{ndim}}
The solutions we obtained in 5-dimensional spacetime can be easily
extended to $(n+1)$-dimensions. The resulting solutions in
(n+1)-dimensions read

\begin{equation}
R(t)=\frac{1}{2}\sqrt{\frac{-1}{C_4}}(e^{(\frac{t}{C_2})}+e^{(-\frac{t}{C_2})}),
 \label{FEND1}
\end{equation}
\begin{equation}
1+a(r)= \frac {1}{1+\frac{C_4}{C^2_2}r^{2}+C_1r^{(\frac
{2\gamma(-n+2)}{n+\gamma-1})}}. \label{FEND2}
\end{equation}

Following these solutions, the energy-momentum tensor which
supports this structure in (n+1)-dimensions will be:

\begin{equation}
\rho=\frac{A}{\pi
GC_2^2}-\frac{Br^{-\frac{(2n-2)(\gamma+1)}{\gamma+n-1}}C_1(\gamma-n+1)}{\pi
GC_4(e^{(\frac{t}{C_2})}+e^{(-\frac{t}{C_2})})^2(\gamma+n-1)},\label{FEND3}
\end{equation}

\begin{equation}
P_r=-\frac{A}{\pi
GC_2^2}-\frac{Br^{-\frac{(2n-2)(\gamma+1)}{\gamma+n-1}}C_1}{\pi
GC_4(e^{(\frac{t}{C_2})}+e^{(-\frac{t}{C_2})})^2},\label{FEND4}
\end{equation}

\begin{equation}
P_t=-\frac{A}{\pi
GC_2^2}+\frac{Br^{-\frac{(2n-2)(\gamma+1)}{\gamma+n-1}}C_1(\gamma-n+3)}{\pi
GC_4(e^{(\frac{t}{C_2})}+e^{(-\frac{t}{C_2})})^2(\gamma+n-1)}.\label{FEND5}
\end{equation}

As a check for the correctness of these solutions one can see that
they reduce to the solutions of \cite{15} for n=3 with proper
definition for the integration constants.

\subsection{Properties of the Solutions\label{field1}}

 Now let us have a look at radial behavior of the solutions. In the paper by
Morris and Thorne \cite{3}, the metric of the wormhole is written
in the form:

\begin{equation}\label{FEND6}
    \mathit{\ ds}^{2}= - \mathit{\ d}\,t^{\mathit{2\ }} +
{\displaystyle \frac {\mathit{\ d}\,r^{\mathit{2\ }}}{1 -
{\displaystyle \frac {\mathrm{b}(r)}{r}} }}  + r^{2}\,(\mathit{\
d}\,\theta ^{\mathit{2\ }} + \mathrm{sin}(\theta )^{2}\,\mathit{ \
d}\,\phi ^{\mathit{2\ }}).
\end{equation}
In which $b(r)$ is the shape function and the throat radius
satisfies $b(r_0)=r_0$. If the equation $d(r)=r-b(r)$ has any root
$r_0$ and simultaneously $d(r)>0$ for $r>r_0$ then we will have a
wormhole and $r_0$ gives the throat radius of the wormhole. In our
solutions, with choosing $\gamma=n-1$ the condition for the
existence of wormholes will be
\begin{equation}
\mathrm{d}(r) = r +\alpha r^{3}  + \beta r^{3-n}=0 \     \ {\rm
at} \ \ r=r_0,
 \label{FEND7}
\end{equation}
and
\begin{equation}
\mathrm{d}(r) = r +\alpha r^{3}+ \beta r^{3-n}>0\      \ {\rm
for}\ \ r>r_0, \label{FEND8}
\end{equation}

where $\alpha$ and $\beta$ are constants related to the
integration constants:
\begin{equation}\label{FEND9}
\alpha = \frac{C_4}{C_2^2}, \             \ \beta = C_1.
\end{equation}

\begin{figure}\epsfysize=4cm
{ \epsfbox{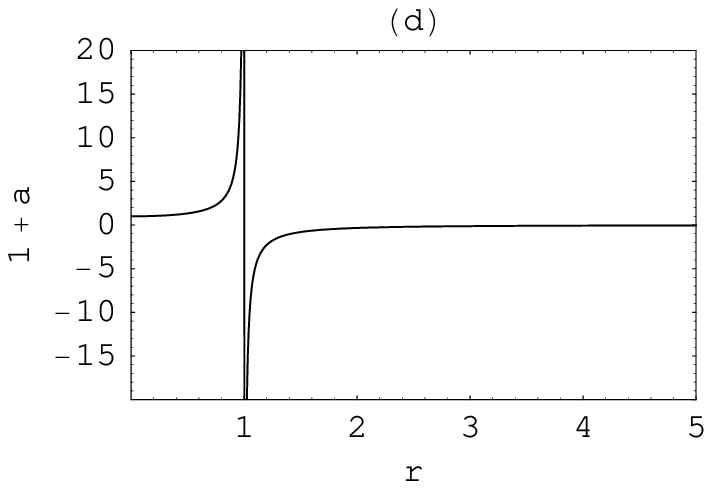}}\epsfysize=4cm { \epsfbox{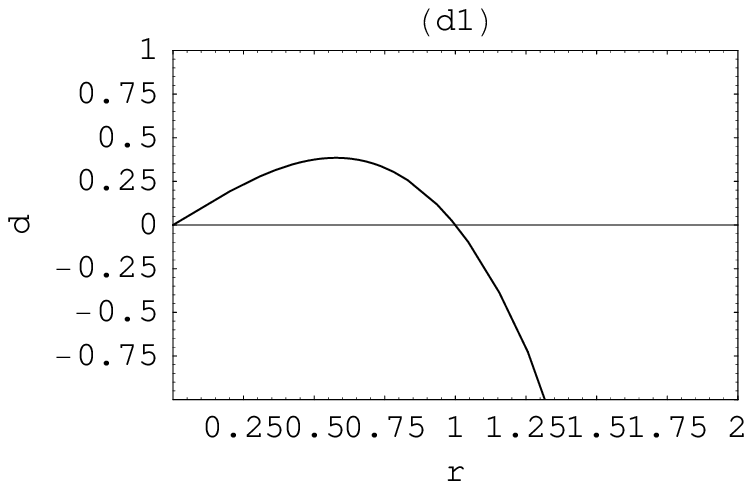}}\epsfysize=4cm
{ \epsfbox{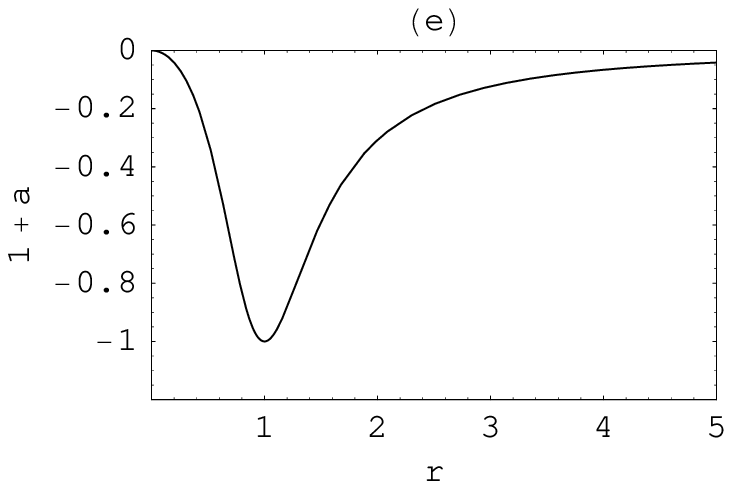}}\epsfysize=4cm { \epsfbox{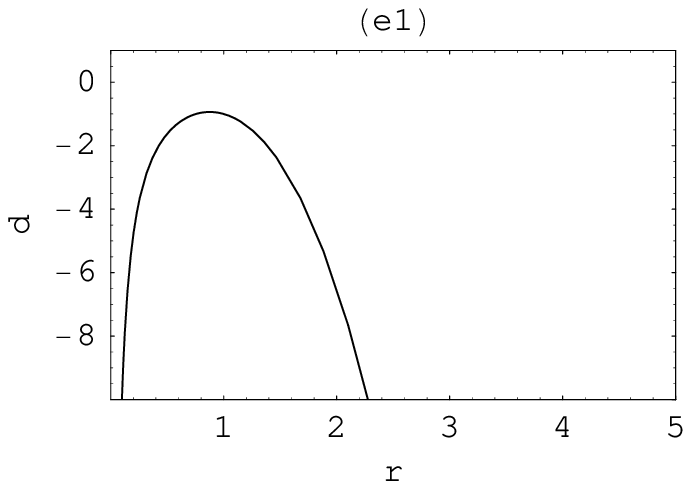}}\epsfysize=4cm
{ \epsfbox{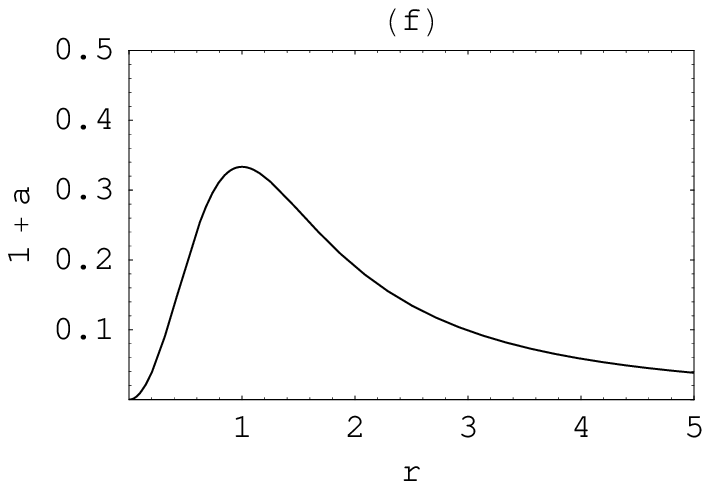}}\epsfysize=4cm { \epsfbox{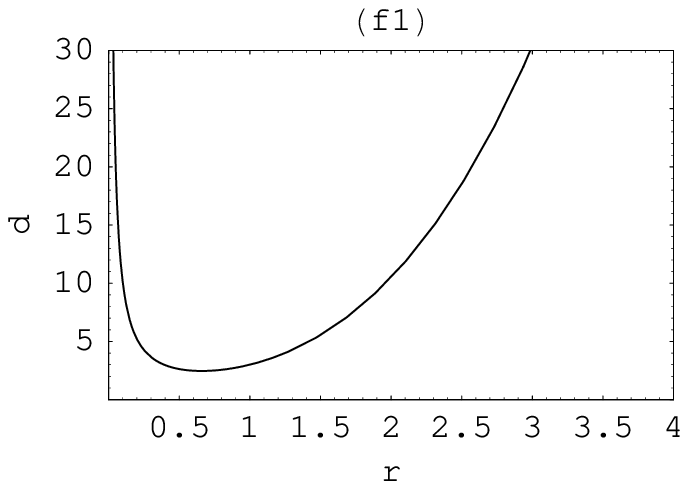}}\epsfysize=4cm
{ \epsfbox{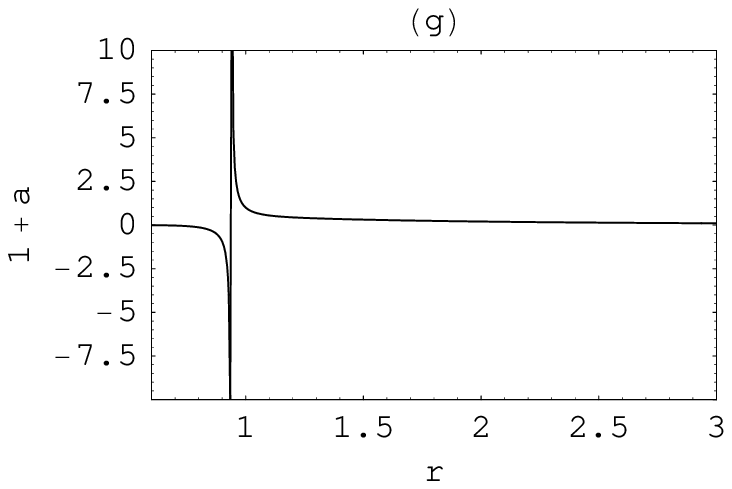}}\epsfysize=4cm { \epsfbox{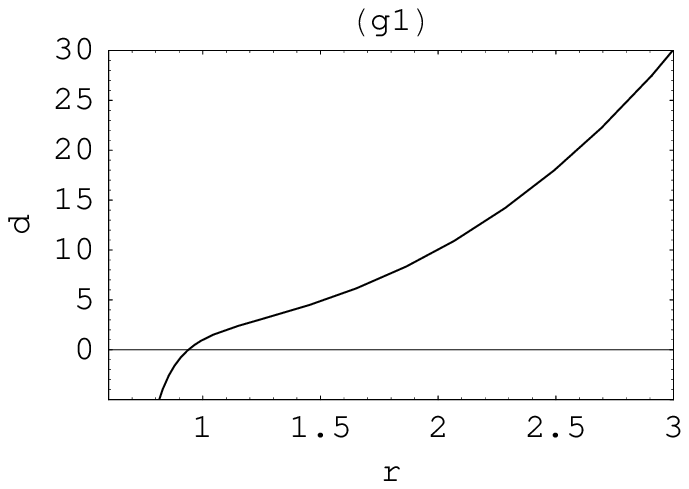}}\caption{The
behavior of $1+a(r)$ and $d(r)$ for $\gamma=3$ and (d)
$\alpha=-1,\beta=0,n=4$, (e) $\alpha=-1,\beta=-1 ,n=4$, (f)
$\alpha=1,\beta=1 ,n=4$, (g) $\gamma=-5,\alpha=1,\beta=-1 ,n=4$}
\label{i2}
\end{figure}

Solving equation (\ref{FEND7}) analytically is not possible except
for special values of n, as presented in \cite{15} for n=3. Then
for investigating the properties of the solutions, we plot $d(r)$
against $r$ in Figure (\ref{i1}-\ref{i2}). We classify the
solutions in the following manner:

\begin{figure}\epsfysize=4cm
{ \epsfbox{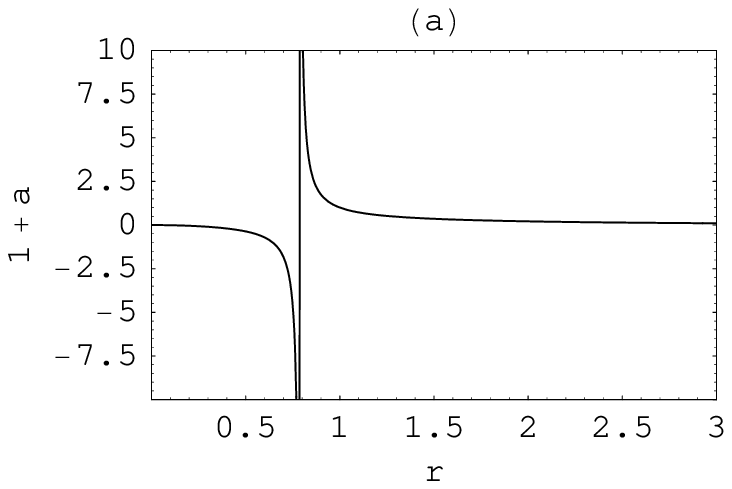}}\epsfysize=4cm { \epsfbox{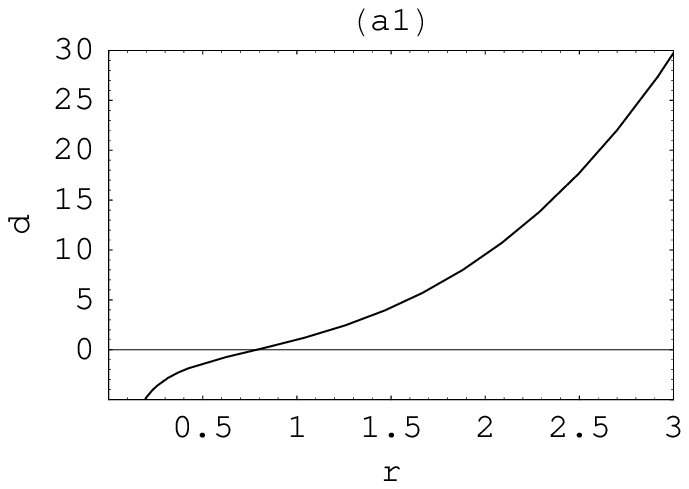}}\epsfysize=4cm
{ \epsfbox{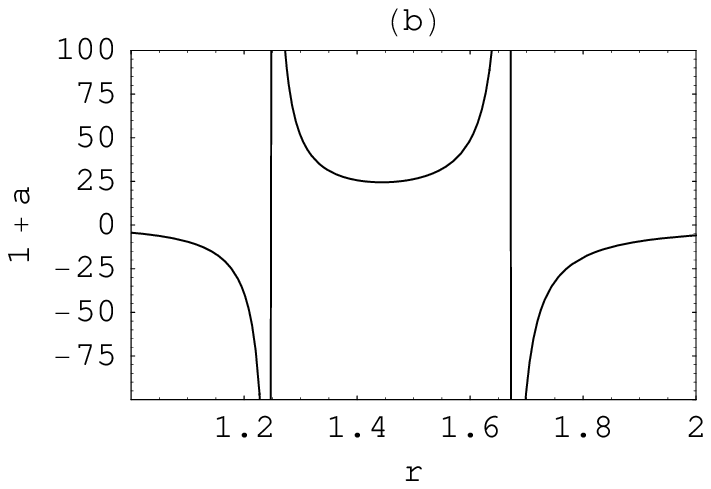}}\epsfysize=4cm { \epsfbox{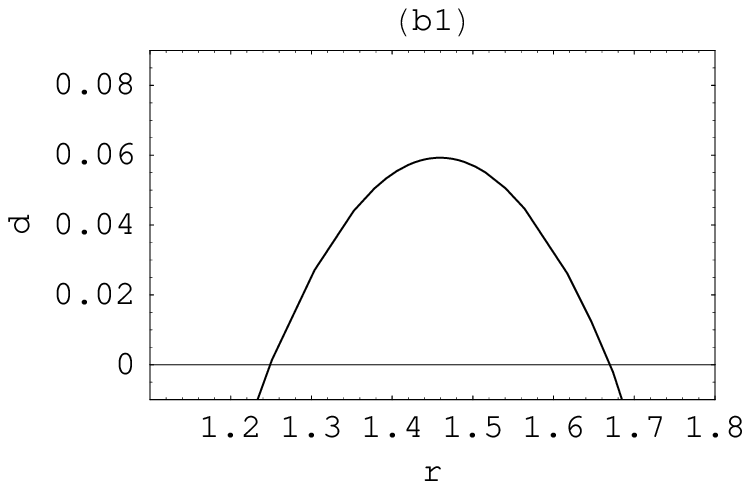}}\epsfysize=4cm
{ \epsfbox{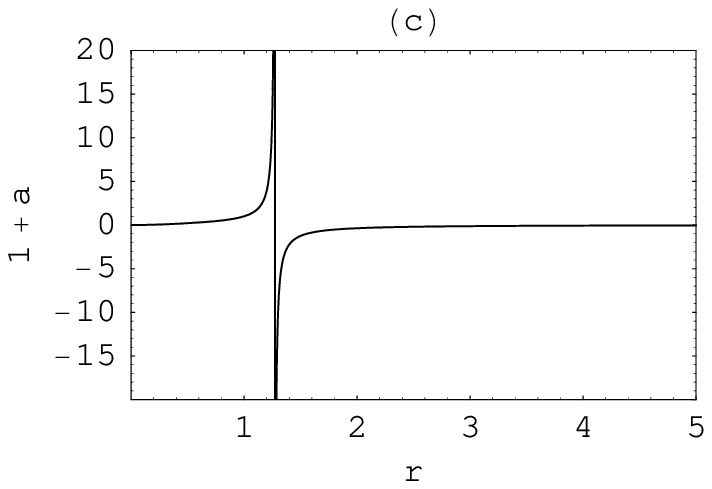}}\epsfysize=4cm { \epsfbox{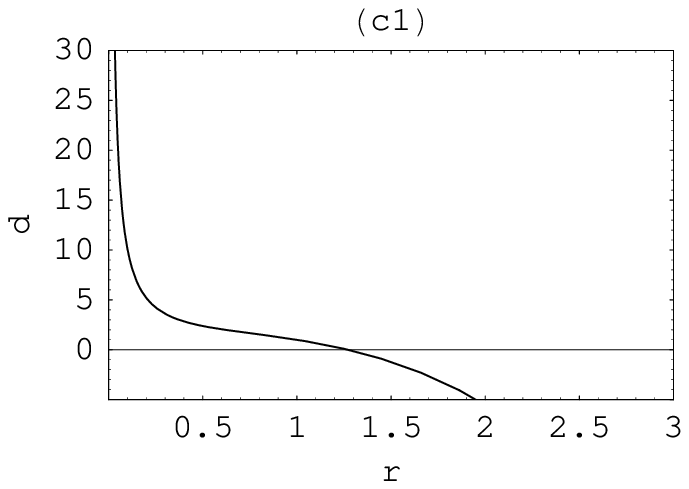}}\caption{The
behavior of $1+a(r)$ and $d(r)$ for $\gamma=3$ and (a)
$\alpha=1,\beta=-1,n=4$, (b) $\alpha=-0.23,\beta=-1 ,n=4$, (c)
$\alpha=-1,\beta=1,n=4$} \label{i1}
\end{figure}

In the case (a), $ \alpha=1$ and $\beta=-1$,  we see from
Fig.\ref{i1} that we have a lower limit on $r$ which corresponds
to the throat radius of the wormhole and we see also that there is
no upper limit on $r$ which reminds us that we have an open
spacetime.

For the choice (b), $ \alpha=-0.23$ and $\beta=-1$, we have lower
and upper limits on $r$. The lower limit corresponds to the throat
radius of the wormhole and the upper limit signifies a closed
spacetime.

In the cases (a) and (b), the Kretschman scalar blows up at $r=0$
but, this point is not included in our physical spacetime with
proper signature.

Case (c), $\alpha=-1$ and $\beta=1$ represents a naked singularity
in a closed cosmological background, because the Kretschman scalar
blows up at the origin ($r=0$).

Case (d), $\beta=0 $ and $\alpha=\pm1 $ leads to $1+a(r)=1/(1\pm
r^2)$, which corresponds to the $k=\mp1$ $FRW$ metric.

Case (e), $\alpha =-1$ and $\beta =-1$, $1+a(r)$ becomes negative
and the metric's signature is not of cosmological interest.

And finally, (f), $\alpha =1$ and $\beta =1$, $1+a(r)$ is regular
everywhere and represent a naked singularity, again but in an open
universe.

We can also have a discussion on the cases which include
$\gamma<0$. For case $\gamma=-5$, from (\ref{FEND2}) we obtain

\begin{equation}
d(r)=r +\alpha r^{3}  + \beta r^{
\frac{11n-26}{n-6}}\label{FEND11}.
\end{equation}

This equation with $\alpha=1$ and $\beta=-1$, leads to a wormhole
centered, open universe which is illustrated in Fig.\ref{i2} part
(g).

It is worth having a look at the Ricci scalar ($\cal R$). In
(n+1)-dimensional spacetime $\cal R$ is:

\begin{equation}
{\cal R}
=\frac{n(n+1)}{C_2^{2}}-\frac{Wr^{(\frac{-(2n-2)(1+\gamma)}{\gamma+n-1})}C_1(\gamma-n+1)}
{(e^{(\frac{t}{C_2})}+e^{(-\frac{t}{C_2})})^2C_4(\gamma+n-1)}.\label{FEND10}
\end{equation}

We can see that choosing $C_1=0$ or $\gamma=n-1$, constant
curvature spacetime is retrieved. In the case $C_1=0$, we have a
maximally symmetric de Sitter spacetime. We expect this result
because, we see that the constant part of the Ricci scalar reminds
us a maximally symmetric spacetime curvature.

The wormholes discussed in this paper are traversable. We present
two reasons here. The first reasoning is based on the redshift of
a signal emitted at the comoving coordinate $r_{1}$ and received
by a distant observer. Using the metric (\ref{FE201}) and for a
radial beam, we obtain

\begin{equation} \label{red1}
\frac{dt}{R(t)}=(1+a(r))dr.
\end{equation}

Using this relation for two signals separated by $\tau_{0}$ in
time when emitted (and $\tau$ when detected), we obtain

\begin{equation} \label{red2}
\frac{\tau}{\tau_{0}}=1+z=\frac{R(t_{0})}{R(t_{1})},
\end{equation}
in which $R(t_{0})$ is the scale factor at the time of
observation, and $R(t_{1})$ is the scale factor at the time of
emission. This leads to the exactly same relation as the
cosmological redshift relation which shows that the wormhole does
not introduce extra (local) redshift. Light signals, therefore can
travel to the both sides of the throat and there is no horizon.

The second is based on the geodesic equation, which -for the
metric (\ref{FE201})- leads to
\begin{equation} \label{red3}
\frac{d^{2}r}{d\lambda^{2}}+\frac{1}{2}\frac{a^{\prime}(r)}{1+a(r)}\left(\frac{dr}{d\lambda}\right)^2+2\frac{\dot
R}{R}\frac{dt}{d\lambda}\frac{dr}{d\lambda}=0
\end{equation}
and
\begin{equation} \label{red4}
\frac{d^{2}t}{d\lambda^{2}}+R\dot
R(1+a(r))\left(\frac{dr}{d\lambda}\right)^2=0,
\end{equation}
in which $\lambda$ is an affine parameter along the geodesic.

The first equation has the following first integral

\begin{equation} \label{red5}
\frac{dr}{d\lambda}=\frac{C}{R^2\sqrt{1+a(r)}}.
\end{equation}

Since the proper distant element is $\delta l=R\sqrt{1+a}\delta
r$, we see that there is no radial turning point and any particle
can move in either radial directions at any point near to the
wormhole, which clearly shows that the wormhole is traversable.

Let us have a brief discussion on the observational consequences
of these solutions. We can have a variety of geodesics in a
wormhole spacetime \cite{Riazinasr}. Geodesics can pass through
the throat right to the other part of the spacetime (the parallel
universe), or be deflected back to the same universe. As pointed
out by Cramer et. al. \cite{cramer}, it should be possible -in
principle- to detect wormholes via the gravitational lensing they
cause. This, however, depends on the wormhole being stable, which
is not addressed in the present paper.

\section{Energy-Momentum Tensor and Exoticity Parameter\label{energy}}
The $(n+1)$-dimensional version of diagonal elements of the
energy- momentum tensor are were given in equations
(\ref{FEND3}-\ref{FEND5}).

Some points are interesting to mention about the energy-momentum
tensor. Since we are looking for spherical structures in a
cosmological background, $P_{t}$, $P_{r}$ and $\rho$ should become
almost $r$-independent at large $r$. In particular, for suitable
values of $\gamma$, the solutions approach $P_{r} = P_{t} = -\rho
= constant$ which correspond to dark energy (cosmological
constant). The second point is the conservation equation of the
energy-momentum tensor:

\begin{equation}\label{E5}
\ D_{\mu}T^{\mu \nu} =0
\end{equation}

which leads to
\begin{equation}\label{E6}
\partial_{t}\rho+n\rho+P_{r}+(n-1)P_{t} =0.
\end{equation}
Our solutions satisfy this equation, as expected.

Let us now investigate the exoticity parameter. For an isotropic
fluid, the energy-momentum tensor is in the form
$T_{\nu}^{\mu}={\rm diag}(-\rho,P,...,P)$ and the exoticity
parameter is defined according to

\begin{equation}\label{E7}
\xi=-\frac{\rho+P}{\rho},
\end{equation}
where $\xi > 0$ corresponds to the exotic matter and $\xi < 0$ to
the non-exotic matter. Since we have an anisotropic medium, the
energy-momentum tensor has the form $T_{\nu}^{\mu}={\rm
diag}(-\rho,P_{r},P_{t},...,P_{t})$. We therefore take the average
pressure
\[
\bar P=\frac{1}{n}(P_{r}+(n-1)P_{t}).
\]

Taking this into account, we adopt the following, more general
definition for $\bar P$:
\[
\bar P=\frac{P_{r}+\gamma P_{t}}{\gamma+1},
\]

and modify (\ref{E7}) according to

\begin{equation}\label{E8}
\bar{\xi} = -\frac{\rho+\bar{P}}{\rho}.
\end{equation}
From (\ref{FE10}) we see that the modified exoticity parameter is
everywhere zero and for all cases, which shows that we are on the
border line between exotic and non-exotic matter according to the
criterion. The weak energy condition $(WEC)$ requires

\begin{equation}\label{E10}
T_{\mu \nu}u^{\mu}u^{\nu} \geq0
\end{equation}
for every nonspacelike $u^{\mu}$ which leads to \cite{22}

\begin{equation}\label{E11}
\rho \geq0      , \                  \  \rho+P_{r} \geq ,\    \and
\rho+P_{t}\geq0 .
\end{equation}

 These equations and relations (\ref{FEND3}-\ref{FEND5}) lead to

 \begin{equation}\label{E12}
    \rho \geq 0 \Rightarrow {\displaystyle \frac {A}{G\,\pi \,\mathit{C_2}^{2}}}  -
{\displaystyle \frac {B\,r^{( - \frac {(2\,n - 2)\,(\gamma  + 1)
}{\gamma  + n - 1})}\,\mathit{C_1}\,(\gamma  + 1 -
n)}{\mathit{C_4}\,\pi \,G\,(\gamma  + n - 1) \,(e^{(\frac
{t}{\mathit{C_2}})} + e^{( - \frac {t}{\mathit{C_2}})} )^{2}}}
\geq 0,
\end{equation}

\begin{equation}\label{E13}
 \rho + P_{r} \geq 0 \Rightarrow  - \frac {2\,B\,C_1\,\gamma \,r^{-\frac{(2n-2)(\gamma+1)}{\gamma+n-1}} }{C_4
\pi G (\gamma  + n -
1)(e^{(\frac{t}{C_2})}+e^{(-\frac{t}{C_2})})^2} \geq 0,
\end{equation}

\begin{equation}\label{E14}
\rho + P_{t} \geq 0  \Rightarrow   \frac
{2\,B\,C_1\,r^{-\frac{(2n-2)(\gamma+1)}{\gamma+n-1}} }{C_4 \pi G
(\gamma  + n - 1)(e^{(\frac{t}{C_2})}+e^{(-\frac{t}{C_2})})^2}
\geq 0.
\end{equation}

These relations show that, with the choice $\gamma\geq 0$, we
can't have any wormhole without violating $WEC$, or at least the
border line at which we have the equality relation in (\ref{E13})
and (\ref{E14}). In the case $\gamma <0$ the relations (\ref{E13})
and (\ref{E14}) are both simultaneously satisfied and we have to
investigate the relation (\ref{E12}) in order to see wether $WEC$
is satisfied. It is interesting to note that for particular ranges
of constants we have wormhole solutions which satisfy $WEC$
throughout the spacetime. As an example of these solutions one can
see case (g) in Fig.\ref{i2}.

\section{Summary and conclusion\label{conclusion}}

We introduced time-dependent wormholes in an $(n+1)$-dimensional
expanding cosmological background. We presented an ansatz metric
and a linear relation between the diagonal elements of the
energy-momentum tensor. With these assumptions, we separated and
solved the field equations. Solutions were classified into
different categories with distinct geometries for the central
object. We distinguished two classes of solutions which
represented Lorentzian wormholes with expanding throat in closed
and open universes. We also found two classes of solutions that
contain an intrinsic singularity in open and closed universes. The
other interesting solution was the familiar maximally symmetric de
Sitter spacetimes. The Ricci scalar was calculated and the
constant curvature class of solutions were discussed. The issue of
the traversability  of the wormhole solutions was considerd. We
investigated the energy-momentum tensor required to support these
solutions. Our solutions led to an energy-momentum tensor which
approaches the cosmological constant case far from the wormhole
for $\gamma \geq 0 $. Finally, we investigated the exoticity
parameter. The exoticity parameter for the linear equation of
state is zero everywhere which is border line between exotic and
non-exotic matter. We also checked the weak energy condition and
found that the weak energy condition in the choice $\gamma \geq 0$
is violated except for the border line case. For the case
$\gamma<0$, $WEC$ is satisfied for suitable values of constants.

 \acknowledgments{N.Riazi acknowledges the support of Shiraz University Research Council.}

\end{document}